# The Science Gateway Community Institute's Consulting Services Program: Lessons for Research Software Engineering Organizations

Marlon Pierce
Indiana University
marpierc@iu.edu

Michael Zentner
San Diego Supercomputer Center
mzentner@ucsd.edu

Maytal Dahan
Texas Advanced Computing Center, UT Austin
maytal@tacc.utexas.edu

Sandra Gesing
Discovery Partners Institute
sgesing@uillinois.edu

Claire Stirm
San Diego Supercomputer Center
cstirm@ucsd.edu

Linda Bailey Hayden
Elizabeth City State University
lbhayden@ecsu.edu

## Introduction

The Science Gateways Community Institute (SGCI) [1,2] is an NSF Software Infrastructure for Sustained Innovation (S2I2) funded project that leads and supports the science gateway community. Major activities for SGCI include a) sustainability training, including the Focus Week week-long course designed to help science gateway operators develop sustainability plans, and the Jumpstart virtual short-course [3]; b) usability and user experience consulting [3]; c) a community catalog of science gateways and science gateway software [4]; d) workforce development activities, including a coding institute for students, internship opportunities, and hackathons [5]; e) an annual conference; and f) in-depth technical support for client gateway projects.

The goals of SGCI's Embedded Technical Support component are to help the institute's clients to create new science gateways or to significantly enhance existing science gateways. Examples of the latter include helping to implement major new capabilities and to implement significant usability improvements suggested by SGCI's usability consultants.

The Embedded Technical Support component was managed by Indiana University and involved research software engineers at San Diego Supercomputer Center, Texas Advanced Computing Center, Indiana University, and Purdue University (through 2019). Since 2016, the component has involved 20 research software engineers as consultants and has conducted 59 client consultations.

## How It Worked

Embedded Technical Support is modeled after the XSEDE Extended Collaborative Support Services program [6], with an SGCI consultant assigned to work at 25% effort on a client's project for six months with possible extensions for up to an additional six months. Potential clients applied for this service through the SGCI website. Projects were taken on a first-come, first-served basis for SGCI's first five years; SGCI currently uses a paid-only model where the requesting project includes a budget for the consultants. SGCI partnered with the XSEDE program as a Level 2 service provider, which allowed the two programs to cooperatively allocate consultants. In addition to direct requests, SGCI services could be requested by researchers through the XSEDE allocation process, and SGCI clients were frequently directed to XSEDE to obtain additional cluster and cloud computing resources.

Client surveys showed that the Embedded Technical Support program was well received and effective. According to exit surveys from 32 out of 55 completed projects, on a scale of 1(lowest score) - 6 (highest score), clients' satisfaction with their consultant was 5.62 with a standard deviation of 0.62. Similarly, when asked to rate the consultation's benefit to them on a scale of 1-6, the average score and standard deviation were 5.54 (0.75). When asked how much the SGCI consultant was able to speed up the client's work, the speedup factor average was 6.4; that is, the consultants were able to help the client do work over six times faster than the client would have otherwise been able to do.

## Conclusions for the Research Software Engineering Community

SGCI has compiled both analytic and anecdotal conclusions on the Embedded Technical Support program. The following are a sampling of our findings.

Multiple Services - SGCI's technical consulting program benefited significantly from other SGCI components, particularly Focus Week and Jumpstart sustainability planning, usability or user experience consulting, and the intern program. Embedding student interns into consultation projects with experienced consultants after the students had gone through other programs (such as coding camps and hackathons) was rewarding and beneficial for the student, client, and consultants.

Customer Relationship Management - Using a customer tracking system was essential for managing numerous clients using multiple services and keeping all team members aware of customer status.

Evaluation - Consultation programs need to be regularly evaluated for effectiveness, and the evaluation process itself needs to be periodically reviewed. Having both clients and consultants aware of a client exit survey that is part of NSF reports is useful for establishing early commitment and accountability from both parties. Evaluations should be tied to key performance indicators. It is useful to have evaluation questions that help identify areas that need improvement and potential for growth. For example, clients rated services and consultants highly, but asking clients if they would pay for the service or if the consultation improved their software engineering processes produced a greater range of responses.

Non-Technical Skills - It is worth some effort for the Research Software Engineering community to determine what makes an effective consultant. Beyond technical skills, successful consultants need to have effective communication skills, be well organized (especially if the client is not), effectively manage their time (as they will have limited effort on any one particular consultation), and to enjoy working on a wide range of problems at the expense of depth.  Even effective consultants reported a level of stress in the number and diversity of consultations in which they were involved.

Teamwork - It is often effective to have consultants work as a team on a specific project.  The Indiana University consultants frequently acted in groups of two or three per project, with one consultant acting as the project manager and lead, one acting as primary developer or engineer, and one acting as scientific and scientific computing consultant, as needed. This is still consistent with the 25% staffing model over the six to twelve months of a typical consultation.

Time Allocation - SGCI's first-come, first-served allocation model resulted in a months-long waiting list, even after some clients were redirected to use XSEDE consulting services. The TrustedCI/CICompass consultation model, which is based on an application review process, may have been more appropriate but also would have introduced delays for newly awarded projects.  The Research Software Engineering community should evaluate different consulting models.

Effort - The 25% effort model for consultants is best suited for small and medium-sized projects (such as NSF CSSI elements and frameworks projects).  SGCI would have benefited from additional funding to scale up support for larger projects to have more significant impact.

Work Plans - Template-based work plans were useful for establishing expectations on what would be delivered and what the client's responsibilities were. We also realized though that the work plan also needed a gap analysis questionnaire that was conducted with clients during the initial interviews. The gap analysis helped identify potential problems, such as the level of a client's commitment or insufficient human resources to support the consultation, the client's access to adequate development resources (such as access to the source codes or systems) to support an external consultant, efforts to redevelop already existing technology, or sufficiency of resources needed to put the results of the consultation into production.

## Acknowledgements

The Science Gateways Community Institute is funded by NSF award #1547611.

## References

1. Lawrence, K.A., Zentner, M., Wilkins‑Diehr, N., Wernert, J.A., Pierce, M., Marru, S. and Michael, S., 2015. Science gateways today and tomorrow: positive perspectives of nearly 5000 members of the research community. *Concurrency and Computation: Practice and Experience*, *27*(16), pp.4252-4268.
2. Wilkins-Diehr, N., Zentner, M., Pierce, M., Dahan, M., Lawrence, K., Hayden, L. and Mullinix, N., 2018. The science gateways community institute at two years. In *Proceedings of the Practice and Experience on Advanced Research Computing* (pp. 1-8).
3. Parsons, P., Gesing, S., Stirm, C. and Zentner, M., 2020. SGCI Incubator and its Role in Workforce Development: Lessons Learned from Training, Consultancy, and Building a Community of Community-Builders for Science Gateways. In *Practice and Experience in Advanced Research Computing* (pp. 491-494).
4. Science Gateways Catalog: https://catalog.sciencegateways.org/#/home
5. Nolte, A., Hayden, L.B. and Herbsleb, J.D., 2020. How to support newcomers in scientific hackathons-an action research study on expert mentoring. *Proceedings of the ACM on Human-Computer Interaction*, *4*(CSCW1), pp.1-23.
6. Wilkins-Diehr, N., Sanielevici, S., Alameda, J., Cazes, J., Crosby, L., Pierce, M. and Roskies, R., 2015, March. An overview of the XSEDE extended collaborative support program. In the International *Conference on Supercomputing in Mexico* (pp. 3-13). Springer, Cham.